\documentclass[aps,pra,twocolumn,groupedaddress,showpacs]{revtex4}
\usepackage{graphicx}
\newcommand{\delete}[1]{}

\newcommand{\be}{\begin{equation}}
\newcommand{\ee}{\end{equation}}

\def\beq{\begin{equation}}
\def\eeq{\end{equation}}
\def\bea{\begin{eqnarray}}
\def\eea{\end{eqnarray}}
\def\ba{\begin{array}}
\def\ea{\end{array}}

\begin{document}

%\preprint{}

\title{Attonewton force detection using microspheres in a dual-beam optical trap in high vacuum}
\author{Gambhir Ranjit, David P. Atherton, Jordan H. Stutz, Mark Cunningham, Andrew A. Geraci}
\email[]{ageraci@unr.edu}
%\homepage[]{http://www.physics.unr.edu/glab/}
%\thanks{}
\affiliation{Department of Physics, University of Nevada, Reno, Reno NV, USA}


\date{\today}

\begin{abstract}
We describe the implementation of laser-cooled silica microspheres as force sensors in a dual-beam optical dipole trap in high vacuum. Using this system we have demonstrated trap lifetimes exceeding several days, attonewton force detection capability, and wide tunability in trapping and cooling parameters. Measurements have been performed with charged and neutral beads to calibrate the sensitivity of the detector. This work establishes the suitability of dual beam optical dipole traps for precision force measurement in high vacuum with long averaging times, and enables future applications including the study of gravitational inverse square law violations at short range, Casimir forces, acceleration sensing, and quantum opto-mechanics.

\end{abstract}

\pacs{42.50.Wk,07.10.Cm,07.10.Pz}
%\keywords{}

\maketitle


\section{Introduction}
Micro- and nano-mechanical oscillators have achieved attonewton force sensitivity, enabling the detection of single-electron spins in solids \cite{rugar2} and tests for non-Newtonian gravity at sub-millimeter length scales \cite{stanford08}. Nanotube resonators have recently demonstrated sensitivity well below the aN$/\sqrt{\rm{Hz}}$ level in a cryogenic system \cite{nanotube}. The ultimate sensitivity of such devices has thus far been limited by thermal noise.  The minimum detectable force in the presence of thermal noise scales inversely with the square root of the mechanical quality factor of the oscillator.  The quality factor is typically limited by materials loss including thermo-elastic dissipation and surface imperfections, as well as clamping loss.  By optically levitating a mechanical oscillator in a high vacuum environment, excellent decoupling is achieved, leading to sub-aN sensitivity even in a room temperature environment \cite{levreview}. Such sensitivity enables new searches for gravitational inverse square law violations at short range \cite{beadprl}, tests of Casimir forces in new regimes \cite{beadprl}, new methods for the detection of gravitational waves \cite{GWprl}, as well as electromagnetic and inertial sensing \cite{levreview}.

Levitated dielectric objects have also been identified as promising candidates for ground state cooling \cite{changsphere,virus}, tests of quantum phenomena in meso-scale systems \cite{changsphere,virus,oriol}, precision interferometry \cite{nimmrichter,hartandy}, and hybrid quantum systems coupled to cold atoms \cite{ranjitpra}.  While the first optical trapping and manipulation of microscopic dielectric particles in vacuum was reported in the 1970s by Ashkin and coworkers \cite{ashkin1,ashkin2,ashkin3}, several recent experiments have revitalized this field, involving mK feedback cooling of dielectric spheres in a dual beam optical trap \cite{raizen}, parametric feedback cooling of nanospheres in an optical tweezer in high vacuum \cite{rochester}, cavity cooling of a nanosphere in an optical cavity trap \cite{aspelmeyercavity}, searches for millicharged particles in an optical levitation trap \cite{millicharge}, and trapping and cavity cooling of a nanoparticle in a combined optical/ion trap in high vacuum \cite{barkeriontrap}.

In this paper, we report the use of laser-cooled silica microspheres as force sensors in a dual-beam optical dipole trap in high vacuum.
A significant challenge in the realization of optically trapped dielectric particles at low pressure has to do with the ability to stabilize the particle while pumping through the regime of intermediate vacuum \cite{ashkin3,aspelmeyercavity}. We have identified a set of trapping and laser-cooling conditions that provide a robust method to pump beads through this transition region between diffusive and ballistic collisions with the surrounding gas molecules. Previous work in a dual-beam dipole trap had achieved high-vacuum lifetimes of order $\sim$ 1 hour \cite{raizen}. Here we have regularly attained trap lifetimes of several days, limited only by applied perturbations which resulted in loss of the particle. In contrast to Ashkin-type single beam levitation traps \cite{ashkin3, millicharge} where the scattering force from a vertically oriented laser beam balances the gravitational acceleration $g$ due to the Earth, the dual beam trap affords a wider tunability of trap parameters. For example, the position of the trapped particle can be made less dependent on the trapping laser intensity.

The  minimum force sensitivity for a harmonic oscillator in thermal equilibrium with a bath at temperature $T$ is
\begin{equation} \label{Fmin1}
F_{\rm{min}}=\sqrt{\frac{4 k_{B} T b k}{\omega_{0}Q}}
\end{equation}
where $b$ is the measurement bandwidth, $k$ is the spring constant of the oscillator, $k_{B}$ is Boltzmann's constant, $w_{0}$ is the resonance frequency, and $Q$ is the quality factor.  In the absence of laser cooling, Eq. \ref{Fmin1} can be written for a microsphere as $F_{\rm{min}}=\sqrt{4 k_{B} T m\Gamma_M b}$
where $\Gamma_M=16P/(\pi\rho v r)$ is the damping coefficient of the surrounding gas, $v$ is the mean speed of the gas, $\textit{m}$ is the mass of the sphere, $\textit{r}$ is the radius, $P$ is the pressure, and $\rho$ is the density of the object.  For a sphere cooled with laser feedback cooling, the temperature in Eq. \ref{Fmin1} becomes $T_{\rm{eff}}$ and the damping rate $\Gamma_{\rm{eff}}$ includes the effect of the cooling laser.
%It is predicted that optically trapped micro-sphere oscillators can attain $Q\sim 10^{12}$ in ultra high vacuum.  With such a high $Q$ and our typical radial trap frequency of $kHz$ the ring down time would be $\tau = \frac{2Q}{\omega_0} \sim 6\cdot 10^8 $ seconds.  Since such a high $Q$ is unacceptable experimentally we use intensity modulated lasers to optically damp (decrease $Q$) and optically cool the center of mass temperature of the of the micro-sphere simultaneously.  Provided that the feedback damping system reduces $Q$ and $T$ by a similar amount, the end result is very little effect on the force sensitivity \cite{Geraci2010}. The trap frequency $\omega_0$ is proportional to the intensity of the trapping laser.  So, since the silica micro-spheres we use have a non-zero absorbtion and a melting point, there is an upper limit to $\omega_0$.
%The two main parameters we have access to which improve force sensitivity are integration time and vacuum pressure.  Reducing chamber pressure turned out to be a non-trivial engineering feat due to turbulence pumping down the vacuum chamber and radiometric forces due to uneven heating of the sphere withing a certain pressure range.
Using this system we have demonstrated a room temperature force sensitivity of order $200$ aN$/\sqrt{\rm{Hz}}$, comparable to that achieved with cryogenic MEMS resonators in Ref. \cite{stanford08}. The corresponding acceleration sensitivity is $\sim 700$ $\mu$g $/\sqrt{\rm{Hz}}$, limited by laser noise. Time-averaged measurements with aN sensitivity have been performed. Measurements with known electric fields applied to charged and neutral beads have been performed to calibrate the detection method. This work establishes the feasibility of versatile, robust, dual beam optical dipole traps for precision force measurement in high vacuum with long averaging times, enabling the studies of gravitational inverse square law violations at short range, Casimir forces, acceleration sensing, and quantum opto-mechanics.

\section{Experimental Setup}
\begin{figure}
\includegraphics[width=0.8\linewidth]{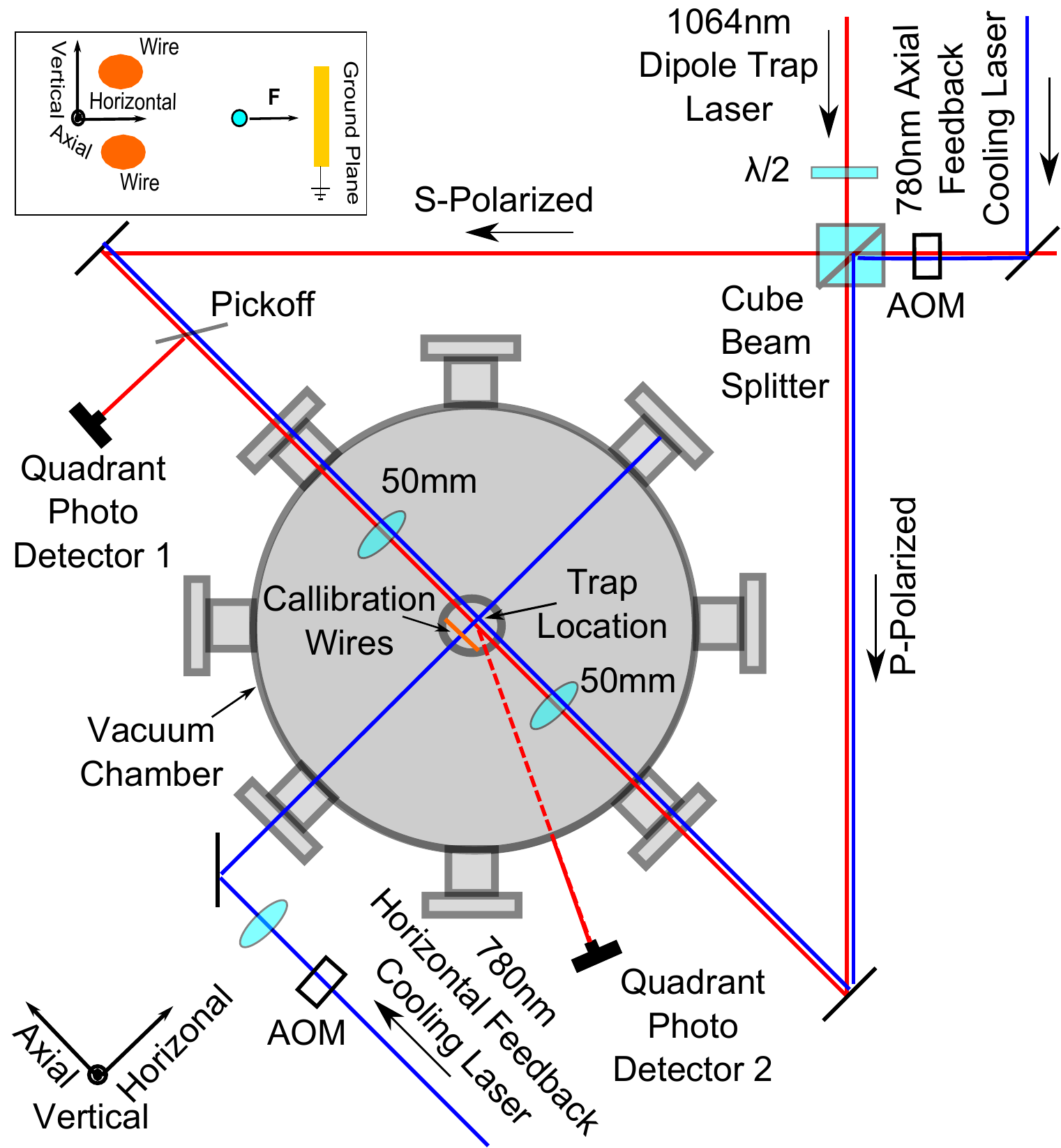}
\caption{\label{fig:experiment}(Color Online) The optical dipole trap is created by focusing two orthogonally polarized laser beams to nearly the same spatial location using lenses with $50$ mm focal length inside a vacuum chamber. $780$ nm laser light is used to provide active feedback cooling to stabilize the trapped beads in vacuum and provide optical damping. (inset) A set of calibration electrodes is used to conduct force measurements with a known applied electric field at the location of the trap.}
\end{figure}

\begin{figure}
\includegraphics[width=\linewidth]{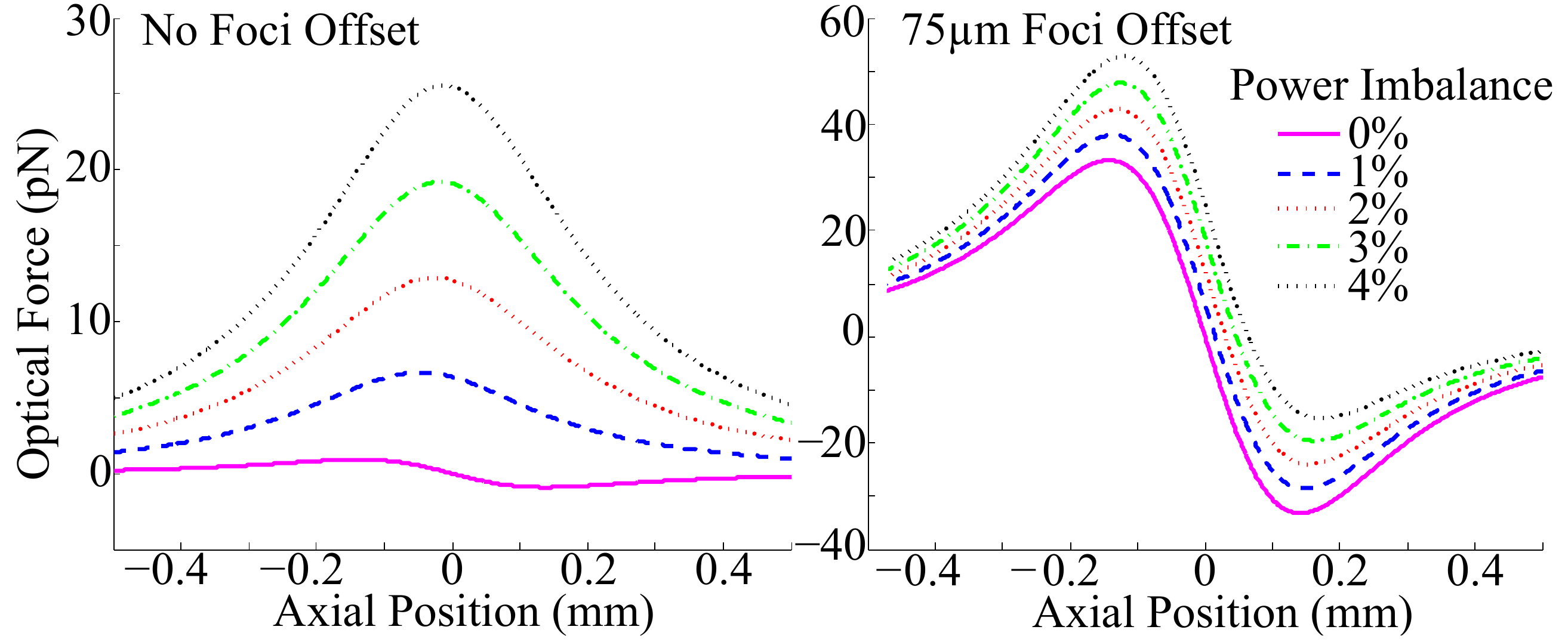}
\caption{\label{fig:offsetFoci}(Color Online) Optical force along the laser axis for the dual beam trap with coincident foci (left) and offset foci (right) for various power imbalance between the S- and P- polarized beams. In this calculation the total laser power is $2.2$ W and the waist is $9$ $\mu$m.}
\end{figure}

The experimental setup is illustrated in Figure \ref{fig:experiment}.  A $3$ $\mu$m fused silica sphere is trapped within an optical dipole trap created by focusing two 1064 nm counter-propagating orthogonally polarized beams at the nearly same position in space. The two beams carry roughly half the total laser power. The dipole trap is initially operated with a total power of $2.2$ W and a waist size of approximately $8$ $\mu$m. The beam foci are offset axially by $75$ $\mu$m to allow for greater imbalance between the incident power of the S- and P- polarized beams. Figure \ref{fig:offsetFoci} shows the optical scattering and gradient force on a bead along the axial direction, as calculated with Lorenz-Mie theory \cite{Miecode}. For zero foci offset, the power balance must be maintained below the 1 percent level to obtain a stable trap. For larger focal separation this requirement is greatly relaxed.  A PZT driven mirror allows in-situ adjustment of the transverse alignment of the beams. To load the trap, an in-vacuum PZT is used to vibrate a glass substrate above the trap center which has beads deposited on it. The trap is typically loaded under low-vacuum of approximately $5-10$ Torr of N$_2$ gas.

The $3D$ position of the micro-sphere is measured using two separate quadrant photo-detectors (QPDs). For active feedback stabilization, the position signals from the QPDs are phase shifted by $90$ degrees to provide a signal proportional to the bead's instantaneous velocity using either a derivative circuit or phase shifter circuit.  The phase shifted signals were used to modulate the RF amplitude of three acoustic optical modulators (AOMs), which modulate the intensity of a $780$ nm laser beam to provide a velocity-dependent optical damping force in each direction. The feedback light is focused onto the sphere using 200 mm lenses outside of the vacuum chamber in the vertical $(y-)$ and horizontal $(x-)$ directions, and using one of the dipole trap lenses for the axial $(z-)$ direction.

\begin{figure}
\includegraphics[width=\linewidth]{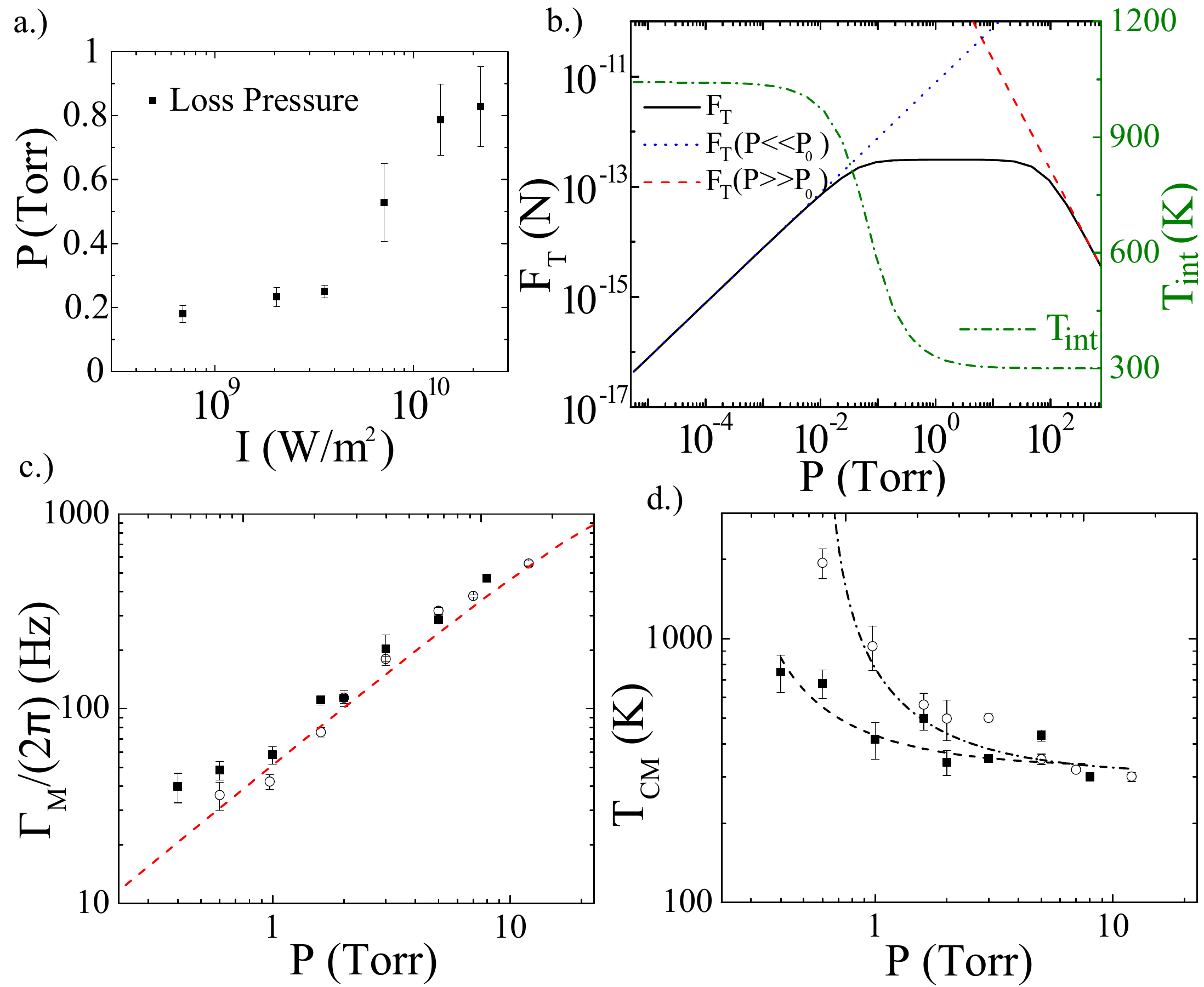}
\caption{\label{fig:beadLoss}(Color Online) (a) Mean pressure at which beads are lost from the trap for various laser trapping intensities, with no laser-feedback cooling applied. Statistics are shown for 30 beads. (b) Calculated internal temperature (right) and photothermal force (left) assuming $\epsilon_2=10^{-6}$ and one percent internal temperature gradient across the bead. (c) Measured damping rate along $x-$ versus gas pressure for two different beads. (dashed line) Calculated damping rate for a bead of diameter $3.0$ $\mu$m in N$_2$ gas. (d) Measured center-of-mass temperature in the $x-$ direction versus pressure. For data shown in (b-d), $I_{\rm{trap}}=2 \times 10^9$ W/m$^2$.}
\end{figure}

Without the application of the laser feedback cooling, particles are lost from the optical trap as the pressure is pumped from approximately 1 Torr to high vacuum. Figure \ref{fig:beadLoss}a shows the pressure at which microspheres are lost from the trap, as a function of laser intensity. There is a marked increase in the pressure at which beads are lost for laser intensity exceeding approximately $4 \times 10^9$ W/m$^2$, whereas the pressure remains relatively constant below this point. Assuming a conservative trap, the trapping depth determined by the dipole potential is given by $U=\frac{3I_{\rm{trap}}V}{c} \frac{\epsilon_1-1}{\epsilon_1+2}$, where $V$ is the volume of the microsphere and $\epsilon_1$ is the real part of the relative permittivity. For our parameters for a 3 $\mu$m sphere the trap depth is approximately $2.5 \times 10^6$K $(\frac{I_{\rm{trap}}}{10^9 {\rm{W/m}}^2})$. Although this trapping depth increases linearly with intensity, the particle loss effect may have to do with instabilities due to nonconservative scattering forces in the optical trap \cite{nonconserv1,nonconserv2} which are enhanced at higher laser power. We have observed that for poor trapping alignment, these nonconservative forces are able to drive cyclic motion in the trap and under these conditions the particle is subject to loss even at pressures exceeding 1 Torr for similar trapping intensities. By steering the trapping beam overlap using PZT mirrors, these effects can be reduced.

The particle loss in Fig. \ref{fig:beadLoss}a also may have to do with radiometric forces \cite{ashkin3,barkerheating}. The finite optical absorption in the sphere results in an increased surface temperature.  As gas molecules collide with the surface, they carry away larger kinetic energy. Currents of these hot air molecules produced by gradients in the trapping intensity can in principle cause the particles to be kicked from the trap \cite{ashkin3}. %Further studies are in process to understand this effect.
We can parameterize the optical absorption through the imaginary part of the complex permittivity $\epsilon=\epsilon_1+i\epsilon_2$. Bulk silica has approximately $\epsilon_2 = 10^{-7}$ \cite{silicaloss}. We can place an upper bound on $\epsilon_2 < 10^{-6}$ by observing that the trapped sphere does not evaporate under high vacuum conditions at high laser intensity. We expect $\epsilon_2$ to have a value within this range. In Fig. \ref{fig:beadLoss}b we show the expected mean internal temperature $T_{\rm{int}}$ of the bead for $\epsilon_2 =10^{-6}$ as a function of gas pressure.  In this model, at high pressure the sphere is cooled through gas collisions, while at high vacuum conditions the heat is dissipated through blackbody radiation. A lower $\epsilon_2$ results in a lower high-vacuum equilibrium temperature, however the shape of the curve is qualitatively similar. $T_{\rm{int}}$ begins to appreciably rise at pressures around 1 Torr.

Fig. \ref{fig:beadLoss}b also shows the expected dependence of the photothermal force on pressure, following the model of Ref. \cite{photothermal}, where we take into account that the temperature gradient depends on the pressure. There is a region between $\sim 100$ mTorr and $10$ Torr where the magnitude of the photothermal force $F_T$ is independent of pressure. This result is in contrast to the early work shown in Ref. \cite{ashkin3} where a fixed temperature gradient was assumed, resulting in a local maximum at $P=P_0$, where $P_0=\frac{3\eta}{r}\sqrt{\frac{R_gT_{\rm{gas}}}{M}}$. Here $\eta$, $T_{\rm{gas}}$, and $M$ are the N$_2$ viscosity, temperature, and molar mass, respectively, and $R_g = 8.31 \frac{{\rm{J}}}{{\rm{mol \cdot K}}}$ is the gas constant. In the flat region the temperature gradient is increasing as $1/P$ while the photothermal force is proportional to $P$. At sufficiently low pressure (e.g. below $10^{-2}$ Torr) the temperature is set by blackbody radiation and the temperature gradient becomes constant. To estimate the magnitude of $F_T$ we assume a $1 \%$ variation in temperature across the $3$ $\mu$m sphere, due to the intensity gradient of the trapping beams near their waists. The shape of this curve is qualitatively similar for larger or smaller temperature variations within an order of magnitude.  Thus there is generally range of intermediate vacuum pressures over which photothermal forces can be significant.

 Taking into account the heating rate due to non-conservative forces from the scattering force, from radiometric forces, and from laser noise, we estimate the steady state average phonon number in the trap (considering only 1-d for simplicity): $\bar{n} = (\bar{n}_{\rm{th}} \Gamma_M +\Gamma_{sc} )/ (\Gamma_M+\Gamma_{\rm{cool}}-\alpha_{\rm{NC}})$. Here $\Gamma_M$ is the damping rate due to the background gas, $\Gamma_{\rm{cool}}$ is the laser cooling rate, and $\bar{n}_{\rm{th}} \Gamma_M$ is the thermalization rate due to the environmental heat bath of the surrounding gas. The term $\Gamma_{\rm{sc}}$ includes heating from laser noise and momentum diffusion due to photon recoil, while the term $\alpha_{\rm{NC}}$ corresponds to heating due to nonconservative forces from the scattering force and the radiometric force: $\alpha_{NC} \equiv \alpha_{\rm{NC (rad)}}+\alpha_{\rm{NC (trap)}}$. In Fig. \ref{fig:beadLoss}c we show the measured damping rate for motion in the $x-$ direction two different beads, with no laser cooling applied. There is reasonable agreement with the calculated damping rate from N$_2$ gas \cite{knudsen}, shown as a dashed line for a sphere with diameter of $3$ $\mu$m, within experimental uncertainties. Fig. 3d shows the corresponding center-of-mass temperature of the beads motion in the $x$-direction as a function of pressure. With laser cooling turned off, i.e. $\Gamma_{\rm{cool}}=0$, we can estimate the value of $\alpha_{\rm{NC}}$ from the pressure at which the beads are lost, corresponding to the vanishing of the denominator in the expression for $\bar{n}$.  At this point the damping rate from the gas is approximately equal to the heating rate from the radiometric forces and nonconservative scattering forces.  The experimentally inferred heating rate $\alpha_{\rm{NC}}$ typically ranges from $20-50$ Hz. Variance in this value may depend on the absorption coefficient of a particular bead, or on the trap alignment.

To avoid the trapping instabilities, we reduce the laser trap intensity $I_{\rm{trap}}$ to approximately $2 \times 10^9$ W/m$^2$ prior to pumping to high vacuum. In this case laser feedback cooling is able to provide damping needed to stabilize the particle as it is pumped to high vacuum using a turbomolecular pump. The intensity of the feedback light used on the particle is typically $\sim 10^7-10^8$ W/m$^2$, in the $x$-, $y$-, and $z$- directions. The feedback gain is increased until the linewidth of the mechanical resonance is roughly $\sim 400-500$ Hz in the transverse directions and $\sim 300$ Hz in the axial direction. The N$_2$ is slowly removed from the chamber by slowly opening a right angle valve over several minutes, until the the chamber pressure is below $10^{-4}$ Torr. At this point the valve can be opened completely and the base pressure of the chamber of approximately $5 \times 10^{-6}$ Torr is reached.  After high vacuum has been attained, the laser cooling rate can be reduced by more than an order of magnitude, while still allowing the particle to remain trapped for long time periods. It is also possible to turn off the feedback in one direction and still maintain the particle in the trap, due to cross-coupling between feedback channels.  This is suggestive that radiometric forces present at intermediate pressures, not solely the nonconservative scattering forces, play a role in the trap loss mechanism.  $I_{\rm{trap}}$ can be increased by a factor of $\sim 5$ at high vacuum pressures before losing the particle. This allows the trapping frequency to be tuned by more than a factor of 2 {\emph{in-situ}}. The loss at high vacuum with high intensity may be due to nonconservative forces in the trap or internal heating.

%\begin{figure}
%\includegraphics[width=\linewidth]{Figures/Ch2Vert_TEST.pdf}
%\caption{\label{fig:cooling}(Color Online) Position spectra for initial and feedback cooled conditions at 1.6 Torr and for feedback cooling at $5 %\times 10^{-6}$ Torr, in three dimensions. Averaging time is 100 s for all data shown. Also shown is the noise floor after the bead is removed %from the trap.}
%\end{figure}

%\begin{figure}
%\includegraphics[width=\linewidth]{Figures/combinedfixed2.pdf}
%\caption{\label{fig:cooling}(Color Online) Position spectra for initial and feedback cooled conditions at 1.6 Torr and for feedback cooling at $5 %\times 10^{-6}$ Torr, in three dimensions. Averaging time is 100 s for all data shown. Also shown is the noise floor after the bead is removed %from the trap.}
%\end{figure}

%Figure \ref{fig:cooling} shows a typical 3D position spectrum of a bead in the initial state, the cooled state at $1.6$ Torr, and the final cooled state at $5 \times 10^{-6}$ Torr, as well as the background noise when the bead is removed from the trap. In this case the final temperature of the bead is approximately 10 mK and $Q \approx 3$.
Prior to pumping to high vacuum, the temperature as derived from the position spectrum of the beads is largely independent of pressure for sufficiently high pressure, as shown in Fig. \ref{fig:beadLoss}d. We thus assume the bead is in thermal equilibrium with the background gas above $5$ Torr.  This allows us to determine a scale factor to convert the quadrant photodetector voltage into a displacement.  %From this conversion factor we can deduce the force sensitivity of the bead at low vacuum.

\begin{figure}
\includegraphics[width=1.0\linewidth]{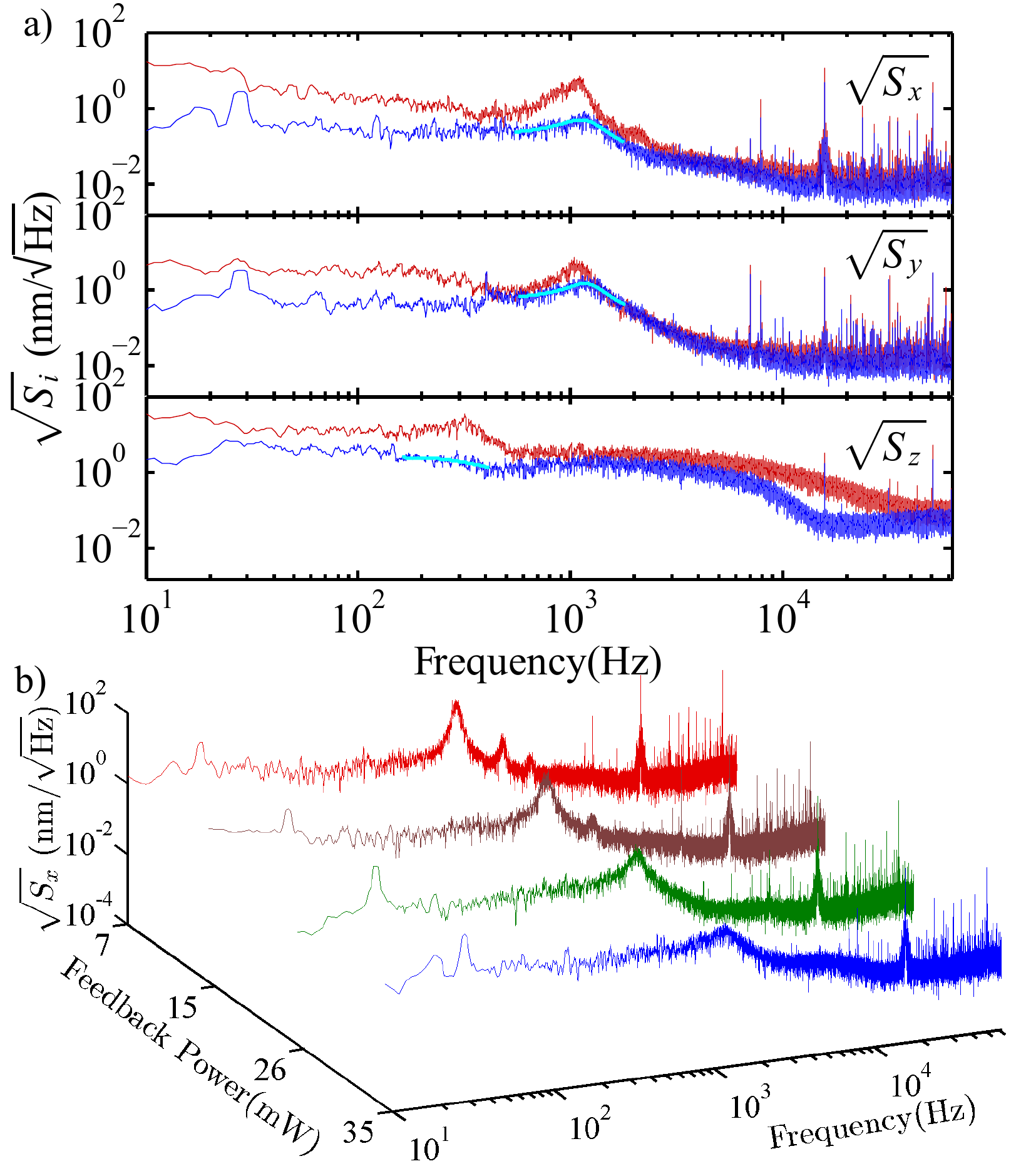}
\caption{\label{fig:waterfall}(Color Online) (a) Typical position spectrum of a bead held at low vacuum of $1.7$ Torr with no feedback cooling applied (red), and at high vacuum of $5 \times 10^{-6}$ Torr with feedback cooling applied (blue). (b) $x$- spectrum of a bead at $5 \times 10^{-6}$ Torr with varying feedback cooling rates.}
\end{figure}

Figure \ref{fig:waterfall}a shows a typical 3-D position spectrum of a bead held at low vacuum of $1.7$ Torr with no feedback cooling applied, and a spectrum at high vacuum of $5 \times 10^{-6}$ Torr with feedback cooling. The transverse modes $(x-,y-)$ are observed with frequencies of $(1073,1081)$ Hz respectively, and the axial $(z-)$ frequency is 312 Hz.  The peaks are slightly shifted when feedback cooling is applied due to the optical spring effect that occurs if the feedback phase is not precisely 90 degrees. Under high vacuum conditions with feedback cooling applied, using a Lorentzian fit we attain effective temperatures of $10\pm3$ K, $55\pm9$ K, and $12\pm2$ K in the $x$-,$y$-,and $z$- directions respectively, with corresponding damping rates of $454 \pm 29$ Hz, $448 \pm 16$ Hz, and $340 \pm 120$ Hz.  The force sensitivity in the $x-$direction corresponds to $S_{F,x}^{1/2}=217\pm48$ aN$/\sqrt{\rm{Hz}}$, with the error dominated by the uncertainties in the particle size and the displacement-to-voltage scaling factor for the quadrant photodetector. The lowest attainable temperature appears to be limited by noise in the trapping laser. The expected sensitivity at this pressure would be approximately $10^2$ times lower in the absence of laser noise and cross-talk between feedback channels.  Fig. \ref{fig:waterfall}b shows the $x$- spectrum of a bead at $5 \times 10^{-6}$ Torr with varying feedback cooling rates.  The beads remain trapped at high vacuum using feedback damping rates less than $10 \%$ of the values needed while evacuating the chamber.

%The beads can be kept at high vacuum pressure using feedback damping rates less than $10 \%$ of the values needed while evacuating the chamber. It is also possible to turn off the feedback in one direction and still maintain the particle in the trap, due to cross-coupling between feedback channels.  $I_{\rm{trap}}$ can be increased by a factor of $\sim 5$ at high vacuum pressures before losing the particle. This allows the trapping frequency to be tuned by more than a factor of 2 {\emph{in-situ}}. The loss at high vacuum at high laser intensity may be due to nonconservative forces in the trap or internal heating.

\begin{figure}
\includegraphics[width=1.0\linewidth]{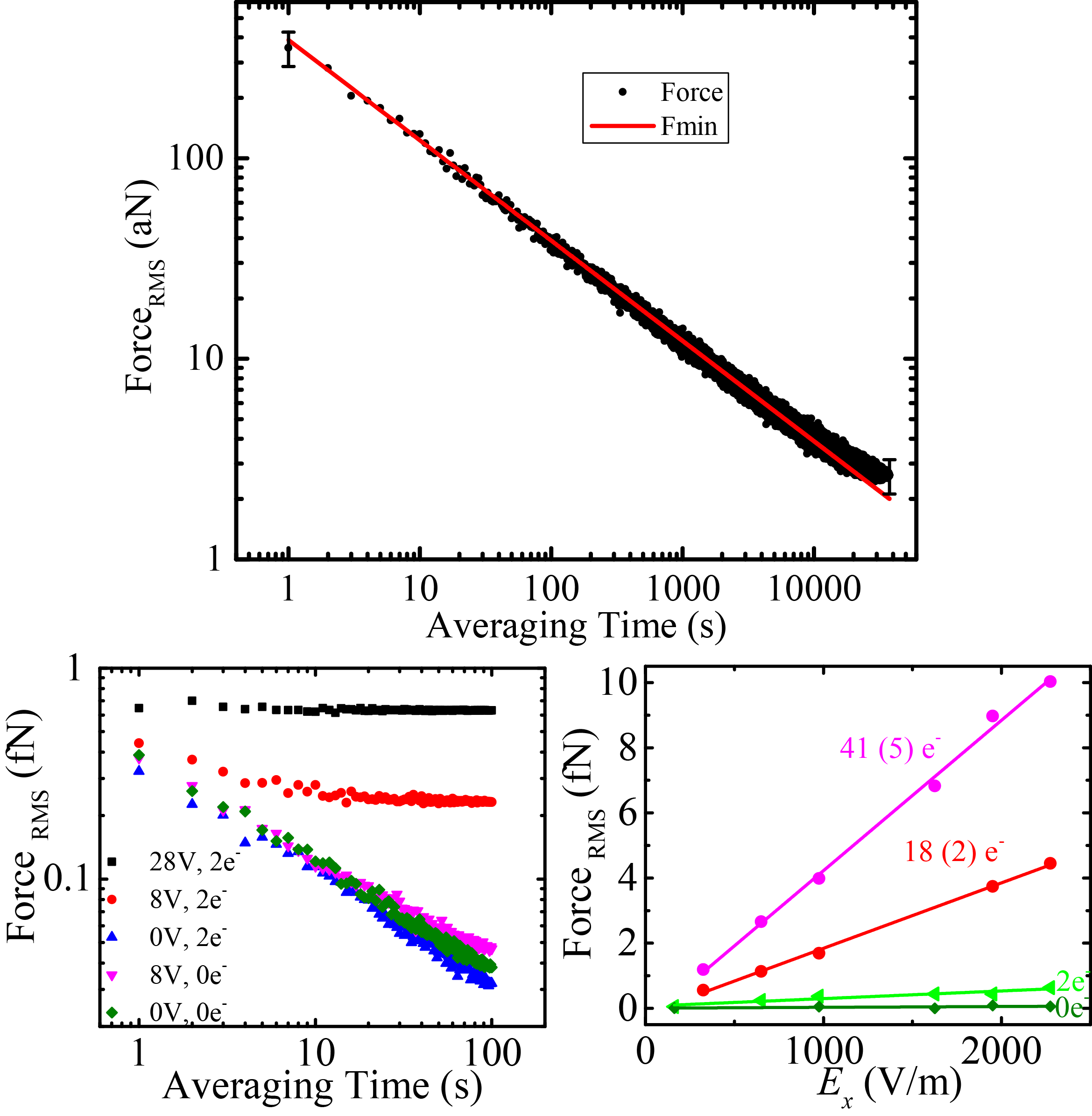}
\caption{\label{fig:thnoise}(Color Online) (a) On-resonance horizontal (x-) force on the bead at $5 \times 10^{-6}$ Torr, versus averaging time, with no applied forces. (b) Force on a charged bead ($2e$) and neutral bead ($0e$) versus averaging time for varying applied driving voltage $V_{AC}$. (c) Force on beads of varying charge (neutral, $2e$, $18\pm2e$, and $41\pm5e$) versus applied driving voltage $V_{AC}$ and corresponding electric field with $100$ s averaging time. Data are shown at $V_{AC}=2,4,8,12,20,24,28$ V.}
\end{figure}

\section{Force Measurements}
To measure an applied force we have included a set of electrodes in the vacuum system to produce a known electric field at the position of the bead.  We perform measurements in the $x-$ direction. In the absence of an applied force, we expect the signal due to thermal noise to average down as the square-root of the measurement bandwidth. This behavior is shown in Fig. \ref{fig:thnoise} for averaging times exceeding $10$ hours.  Force sensitivity at the level of $\sim 2$ aN is achievable at this timescale.
Figure \ref{fig:experiment} depicts the wire configuration we used to apply an electric force on the sphere. %A sinusoidal voltage is applied using a Stanford Research Systems DS345 signal generator. This signal is amplified by a high-frequency amplifier (A.A.Labs A-303) and sent to the calibration electrodes.
We apply a sinusoidal AC voltage ranging from $V_{AC} \sim 2-28$ V to the two wires, corresponding to electric fields up to $2300$ V/m.  We find that approximately $80\%$ of beads are trapped with a non-zero electric charge. Those beads which are neutral after trapping remain neutral indefinitely over the timescale of our measurements, which we have extended over several days.  %These neutral beads are necessary for precision measurement of Casimir forces or gravitational forces.
The beads which are charged tend to retain the same value of charge. By implementing suitable UV light source, control of the bead charge should be possible \cite{millicharge}. In earlier work at low vacuum (1.7 Torr), we have found that by applying UV light in the vacuum chamber the bead charge can be reduced {\it{in-situ}} \cite{jordanthesis}. %It is useful to also perform measurements with charged beads, to demonstrate the ability to measure a larger known applied force, and to calibrate the force sensitivity if the number of electrons is known.

%The electric field from the wires is primarily directed along the horizontal axis at the trap location.  %A sphere with polarizability $\alpha$ and charge $q$ experience a force ${\bf{F}}= \frac{1}{2}\alpha {\bf{\nabla}} E^2 + q{\bf{E}}$ when interacting with an electric field ${\bf{E}}$.
In Fig. \ref{fig:thnoise}b we show the force on a charged bead and neutral bead as a function of averaging time, for different values of applied electric field. When the applied field is turned off, the noise floor averages down with the inverse square root of averaging time as expected. These measurements are performed near the trap resonance frequency $\approx$ 1.1 kHz. Measurements are also taken off resonance at 7 kHz, where the mechanical response of the bead is significantly reduced, and without a bead in the trap to determine the electronic background noise. We find the electronic noise is not significant over the 100 s timescale for driving voltages $V_{AC}$ up to 8 V.
In Fig. \ref{fig:thnoise}c we show the force versus applied electric field for several different beads with differing charges. The sign of the charge is determined by measuring the phase of the motion with respect to the applied voltage. Measurements with beads having a known small number of electrons (e.g. 1 or 2) provide an independent calibration of the force sensitivity.

\section{Discussion}
%We have demonstrated force sensitivity at the aN level using microspheres trapped in a dual-beam optical dipole trap at high vacuum, with trapping lifetimes exceeding days.
We have determined a set of trapping and cooling parameters which permit pumping through the intermediate vacuum transition where trap instabilities are known to be present. At high vacuum we have demonstrated force sensitivity at the aN level, with trapping lifetimes exceeding days. We estimate that improved sensitivity of order $10^2$ can be attained with reduced laser noise at similar pressures, for example by using an intensity-stabilized laser in a standing wave trap to reduce the effect of beam pointing fluctuations in the counter-propagating beams. Also nonconservative heating effects due to the scattering force should be reduced in such a trap, due to improved overlap of the counter-propagating lasers. Our method produces trapped beads with zero or negative charge, which remains constant over the trap lifetime at high vacuum.  This system shows promise for precision measurements in ultra-high vacuum requiring long averaging times, including tests of short-range gravitational forces or Casimir forces, as well as experiments on quantum optomechanics, for example where silica beads are held in a dual-beam dipole trap and cooled using a cavity \cite{virus}, with feedback cooling \cite{cavityless}, or by sympathetic cooling with atoms \cite{ranjitpra}.


\section{Acknowledgements}
We are grateful to Melanie Beck and Darren Zuro for experimental assistance at the early stages of this work. We thank T. Li, J. Weinstein, M. Aspelmeyer, and N. Kiesel for useful discussions. This work is supported by grant NSF-PHY 1205994.

%\bibliography{references}

\end{document}